\documentclass{aa}
\pdfoutput=1
\usepackage{graphicx}
\usepackage{txfonts}
\usepackage[]{natbib}
\usepackage{color}

\begin{document}

\title{The Probability Distribution Function of Gas Surface Density in M33 }

\titlerunning{PDF in M33 }

\author{Edvige Corbelli
       \inst{1}
        \and
           Bruce G. Elmegreen
       \inst{2}
           \and
           Jonathan Braine
	\inst{3}
           \and
           David Thilker
	\inst{4}
     }

   \institute{INAF-Osservatorio Astrofisico di Arcetri, Largo E. Fermi, 5,
             50125 Firenze, Italy\\
             \email{edvige@arcetri.astro.it}
             \and
              IBM Research Division, T.J. Watson Research Center, 1101 Kitchawan Road,
              Yorktown Hts., NY 10598\\
              \email{bge@us.ibm.com}
	     \and
             Laboratoire d'Astrophysique de Bordeaux, Univ. Bordeaux, CNRS, B18N, 
                   all\'ee Geoffroy Saint-Hilaire, 33615 Pessac, France\\
                   \email{jonathan.braine@u-bordeaux.fr}
	    \and
             Center for Astrophysical Sciences,The Johns Hopkins University, 
             3400 N.Charles Street, Baltimore, MD 21218, USA\\
             \email{dthilker@pha.jhu.edu}
           }

   \date{Received .....; accepted ....}

 \abstract
   {}
   { We examine the interstellar medium (ISM) of M33 to unveil fingerprints of self-gravitating
   gas  clouds across the star-forming disk.  }
   {The probability distribution functions (PDFs) for atomic, molecular, and total gas surface densities
    are determined at a resolution of about 50~pc from available HI and CO  emission line data  
   over regions that share coherent morphological properties and considering cloud samples at different evolutionary 
   stages along the star formation cycle. }
   {Most of the total gas PDFs from the central region to the edge of the star-forming disk
   are well-fitted by log-normal functions whose width decreases radially outwards. Because the HI
   velocity dispersion is approximately constant across the disk, the decrease of the PDF width is consistent  
   with a lower Mach number for the turbulent ISM at large galactocentric radii where a higher fraction 
   of HI is in the warm phase.  The atomic gas  
   is found mostly at face-on column densities below N$_{H}^{lim}$ = 2.5 10$^{21}$~cm$^{-2}$, 
   with small radial variations of N$_{H}^{lim}$.
   The molecular gas PDFs do not show strong deviations from log-normal functions in the
   central region where molecular fractions are high. Here the high pressure and rate of star formation
   shapes the  PDF as a log-normal function  dispersing self-gravitating complexes with intense feedback
   at all column densities that are spatially resolved.
   Power law PDFs for the molecules are found  near and above N$_H^{lim}$,
   in the well defined southern spiral arm and in a continuous dense filament extending  at larger
   galactocentric radii. In the filament nearly half of the molecular gas departs from a log-normal PDF and power laws 
   are also observed in pre-star forming molecular complexes.
   The slope of the power law is between $-1$ and $-2$. This slope, combined with
   maps showing where the different parts of the power law PDFs come from, suggest a
   power-law stratification of density within molecular cloud complexes, which is consistent with the dominance
   of self-gravity.}
  {}

   \maketitle

\section{Introduction}
\label{intro}

The formation and evolution of self-gravitating structures in the interstellar medium
(hereafter ISM) is a key ingredient of star formation.
Stars form in the dense parts of interstellar clouds where self-gravity overwhelms
local gas pressure, causing collapse \citep{mckee07}. The connections between
these dense parts and their surrounding gas can reveal the processes involved. If
the cloud is part of a shell \citep{palmeirim17}, converging flow
\citep{2018PASJ...70S..55W} or spiral arm shock front \citep{elmegreen14}, then the clouds
are likely to have been formed or influenced by the associated pressures, after
which gravity could have made the central parts dense enough for star formation.
If disk gravitational instabilities generate spiral arms and filaments that
fragment into molecular clouds, then even  the birth of these complexes might be
driven by gravity out to the edges of their HI envelopes.

The column density and density where   environmental processes lose their influence
and self-gravity takes over are indicators of the driving pressures and timescales for
cloud evolution. The column density at the threshold for strong gravity in a medium of
pressure $P$ is $\Sigma_{\rm thres}\sim\left(2P/\pi G\right)^{0.5}$, and the dynamical
time for evolution is proportional to the idealized free fall time at density $\rho$,
which is $t_{\rm ff}=\left(32 G\rho/3\pi\right)^{-0.5}$. In interacting galaxies where
pressure is high, the column density and density at the transition should both be high,
making star-forming clouds opaque, fully molecular and quickly evolving. Conversely, in
the far-outer parts of spiral galaxies and in dwarf irregulars where the pressure is
low, the transition should be at a low density, making the timescales long and possibly
placing the first stages of collapse in the atomic medium
\citep[e.g.,][]{michalowski15,meidt16,elmegreen18}.

The best way to determine the transition point from environmental to internal cloud
control is by mapping the dynamical state of the cloud  and its surroundings
 (i.e. which region is collapsing, accreting, or in virial equilibrium,  and 
the type of support against gravity). 
Our galaxy offers the possibility to study individual star forming regions, identify 
dense clumps, filaments, shocks, and map their column densities and velocity fields.
This has been done in several local regions 
\citep[e.g.,][]{deharveng15,figueira17} but it is difficult to do in external galaxies
\citep{hirota11,donovan13,colombo14,tosaki17,baba17,faesi18} because of a lack of
spatial resolution needed to identify high column density clumps.

An alternative to cloud-scale mapping is the use of column density probability
distribution functions (PDFs) from maps of gas emission or absorption. The PDF for gas
in a large-scale map gives the probability distribution of column density averaged over
all of the clouds and intercloud regions. This does not allow the history of individual
star-forming clouds to be determined, but it can give the average column density where
cloud self-gravity begins to dominate over environmental pressure as a driving force. This
is possible because self-gravity gives a cloud an internal density structure that is a
power law with radius, such as $\rho(r)\propto r^{-1.5}$ or $r^{-2}$
\citep[e.g.][]{shu77}, as observed, e.g., by \cite{mueller02}. 
Then the PDF has a
matching power law at high density even if the cloud has  locally a log-normal PDF from
compressible turbulence \citep{elmegreen11}. 
Small scale collapse inside a turbulent cloud also produces a power-law tail   
which extends to lower densities as time proceeds, while turbulent motions broaden the PDF 
\citep{2014ApJ...781...91G}. 
Computer simulations showing this power law
extension are in \cite{klessen00}, \cite{vaz08}, \cite{kritsuk11}, \cite{fed13},
\cite{pan16} and elsewhere.  Log-normal distributions without self-gravity were first
discussed and simulated by \cite{vaz94}, \cite{pvs98}, and \cite{pnj97}. The lower
limit to the power law part of the PDF corresponds approximately to the column density
at the edge of the self-gravitating part of the cloud, and therefore gives the pressure
at that place \citep[e.g.,][]{kainulainen11}. 
 Sensitivity limits, however, are important and should be taken into account  since,
 for example, the total PDF may not curve over at low column density if the detection
limits for gas are close to the apparent PDF peak \citep{alves17}.

PDFs have been extensively studied  and observed in the local interstellar medium using
also extinction or dust emission. In the  Milky Way the resolution  is sufficiently
high to trace the molecular clumps that form individual stars
\citep{kainulainen09,kainulainen11,froebrich10,lombardi10,schneider12,schneider15b}.
\cite{lombardi10} found a log-normal PDF using
extinction maps in a large region around Perseus and Taurus in the solar neighborhood, with a
slight excess over the log-normal above $A_{\rm V}\sim2$ mag. Power law PDF extensions
were also found in dust emission \citep{schneider13,schneider15a,schneider15c,lombardi15} 
and molecular line emission \citep{schneider16} up to $A_{\rm V}\sim100$ mag \citep{schneider15b}.
\cite{schneider12} measured the PDFs of dust emission in the Rosette molecular cloud
and found it became a power law above $A_{\rm V}=9$ mag in the central regions and
above 4 mag in other regions; it was a log-normal far from the Rosette nebula.  
It is still not clear, however,  whether non star-forming clouds have  power-law extensions. 
Although in most external galaxies we cannot yet resolve the cores of individual
star-forming clumps, we may still trace where gravity takes over ISM pressure on
a larger scale, and it can be useful to  compare the PDFs in various environments, such as
galaxy centers, spiral arms and interarms, and as a function of galactocentric
distance and cloud type.  

M33 is a good candidate to examine for these questions because it is the nearest isolated spiral
galaxy. Recent surveys of the  ISM throughout the whole star-forming 
disk of M33 have mapped the atomic and molecular gas phases with high resolution
\citep{2012A&A...542A.108G,druard14,2014A&A...572A..23C}. The H$\alpha$
and GALEX ultraviolet data complement the Spitzer mid-infrared survey to trace
star formation from the embedded to the exposed Young Stellar Clusters 
\citep{2009A&A...493..453V,2011A&A...534A..96S}. The possible association between 566  Giant
Molecular Clouds (hereafter GMCs)
and 600 Young Stellar Cluster Candidates (hereafter YSCCs) has helped in defining the
duration of the lifecycle of
GMCs in M33 \citep{2017A&A...601A.146C}.  Still, it is unclear what drives the transition
from inactive gas to a star-forming cloud. Also, if stellar feedback disrupts these clouds, it
is  of interest to see  if this disruption affects only the small scale structure
of the ISM or if it also contributes to the global disk morphology
\citep{2000ApJ...540..797W}. The presence of turbulence in the ISM of M33 has been
investigated via power spectra using optical or far-infrared images
\citep{2003ApJ...593..333E,2012A&A...539A..67C}.
A recent numerical simulation of the large scale structure of the M33 disk
\citep{2018MNRAS.tmp.1241D} points out the role of both stars and gas in
triggering the formation of high-density filaments by gravitational instabilities.
At the same time, the presence of very low density cavities in the interarm
regions underlines the need for a high level of feedback (e.g., $\sim10$\% of the
total energy of massive stars) to deposit turbulent energy. 
The purpose of this paper is to understand
the role of turbulence and gravity in triggering the formation of GMCs and
star formation throughout the disk of M33.
Due to the strong evidence of non-negligible turbulent motion in the M33 disk we shall
interpret the log-normal PDFs that we detect in the ISM of this galaxy as due to
turbulence although  there might be other features associated to clouds that can
produce log-normal column density distributions \citep{2010MNRAS.408.1089T}.

The analysis of the PDF for the molecular gas in M33 by \citet{druard14} has shown  
deviations from a log-normal shape for gas column density above $\sim 1.7\times 10^{21}$~cm$^{-2}$.
The authors attributed the deviations of a galaxy-wide PDF to self-gravity, but
do not draw any definitive conclusion on the cloud scale affected by gravitational
contraction. The coarse spatial resolution of CO observations in external galaxies in fact does not
allow analysis of the details of individual clouds and hence to distinguish a free-fall contraction of the
whole GMC complex from self-gravity of clumps and cores inside them. The publication of a complete
GMC catalogue in the star forming disk of M33 and of further analysis of the GMC properties, that
followed the complete census of molecular gas presented by \citet{druard14}, open the possibility to
investigate the role of gravity in different types of GMCs which are likely GMCs at different evolutionary stages. 
Furthermore, making use of high angular resolution HI data products presented by \citet{2014A&A...572A..23C} 
we can examine the PDFs of atomic and total gas .   For the molecular component we extend the analysis of 
\cite{druard14} to specific disk areas or cloud types where self-gravity  or turbulence plays a major role, and 
determine if the PDFs depend on  star formation rate or 
galactocentric distance and what this implies   on the resolved scale of $\sim50$ pc.  This is particularly
relevant for M33 given  that both  disk  gravitational instabilities 
and high levels of  stellar feedbacks play a role in shaping the  atomic gas distribution in the disk
\citep{2018MNRAS.tmp.1241D}.

In what follows, the data and cloud types are summarized in Section \ref{data},
the PDFs and power law slopes are presented in Section \ref{pdf}.
Results are discussed in Section \ref{models} and summarized in Section \ref{sum}.

\section{The data}
\label{data}

In analyzing the data of M33, we assume a distance of 840~kpc  \citep{1991ApJ...372..455F,2013ApJ...773...69G}
which implies a linear scale of 4.1~pc per arcsec and 0.24~kpc per arcmin. As
position angle and inclination of the gaseous disk we use PA=20~$^\circ$ and $i$=54~$^\circ$
based on the averages for a tilted ring model fitted to the 21-cm velocity field
at R$<$7~kpc \citep{2014A&A...572A..23C}. The PDF is analyzed for face-on values of  
H-atoms column densities.

\subsection{The neutral atomic gas  traced by HI  21-cm line emission }

To analyze the distribution of  the neutral atomic gas, we use the high resolution
Very Large Array (VLA) and the Green Bank Telescope (GBT) observations of the 21-cm HI line 
emission described by \citet{2014A&A...572A..23C}.
We generally choose an angular resolution that is less than the limit in the
VLA+GBT survey in order to trace more sensitively the faint diffuse emission, and a 
spectral  resolution of 1.29~km~s$^{-1}$. The moment task in MIRIAD has been used to
generate the moment maps. 
The moment-0  is the integrated intensity map of the 21-cm line, 
the moment-1, or mean velocity, is the map of the weighted mean velocity along the line of sight, 
and the moment-2, or   velocity
dispersion, is the map of the intensity-weighted mean deviation of the velocity along the line of sight.
For each pixel $i$ the values of integrated intensity, $I_i$,
mean velocity, $V_i$, and velocity dispersion, $\sigma_{21,i}$,  are computed using $S_{i,j}$, the 21-cm line
intensity in the velocity channel centered on $v_j$,  as follows: 
  
\begin{eqnarray}
 I_i = \sum_j S_{i,j} \qquad \qquad V_i =  \frac {\displaystyle \sum_j S_{i,j}\times v_j}  {\displaystyle \sum_j S_{i,j}}  
 \end{eqnarray}
 \begin{eqnarray}
\sigma_{21,i} = \sqrt{{\displaystyle\sum_j S_{i,j}\times (v_j-V_i)^2 \over \displaystyle\sum_j S_{i,j} } }  
\end{eqnarray}

\noindent
where the sum  extends over all spectral channels with detectable emission.
The images we  use in this paper mostly have a 10~arcsec (41~pc) restored 
beam size and are masked to mitigate contributions from foreground Milky Way emission.  
The pixel size is 4~arcsec, smaller than the beam size.
To check the high column density PDF we also use images at 5~arcsec spatial resolution
with pixel size of 2~arcsec.
The radial trend of the HI line width is inferred from the moment-2
map at a spatial resolution of 10~arcsec.

Analysis of the moment-0 map at 10~arcsec resolution indicates a noise standard
deviation $\sigma_{HI}$=0.0049~Jy/beam~km~s$^{-1}$ = 5.4$ \times 10^{19}$~H cm$^{-2}$
on a pixel by pixel basis . We
consider only pixels with a flux above $3\sigma_{HI}$,
and trace the statistical analysis down to this limiting flux value, which corresponds to a face-on
column density of about  $10^{20}$~~H~cm$^{-2}$.

\begin{figure*}
\centerline{
\hspace*{1cm}\includegraphics[width=12cm]{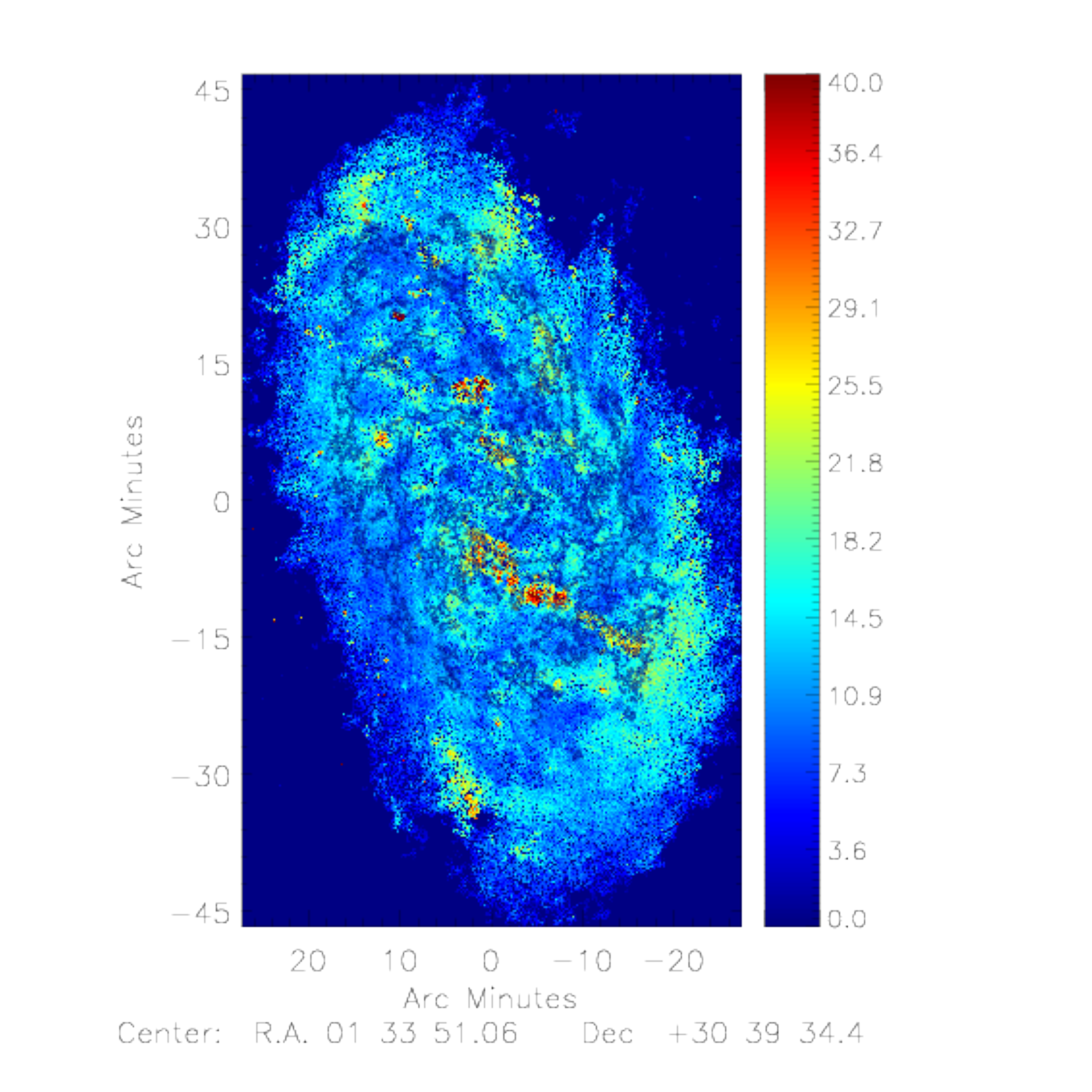}
\hspace*{-3cm}\includegraphics[width=12cm]{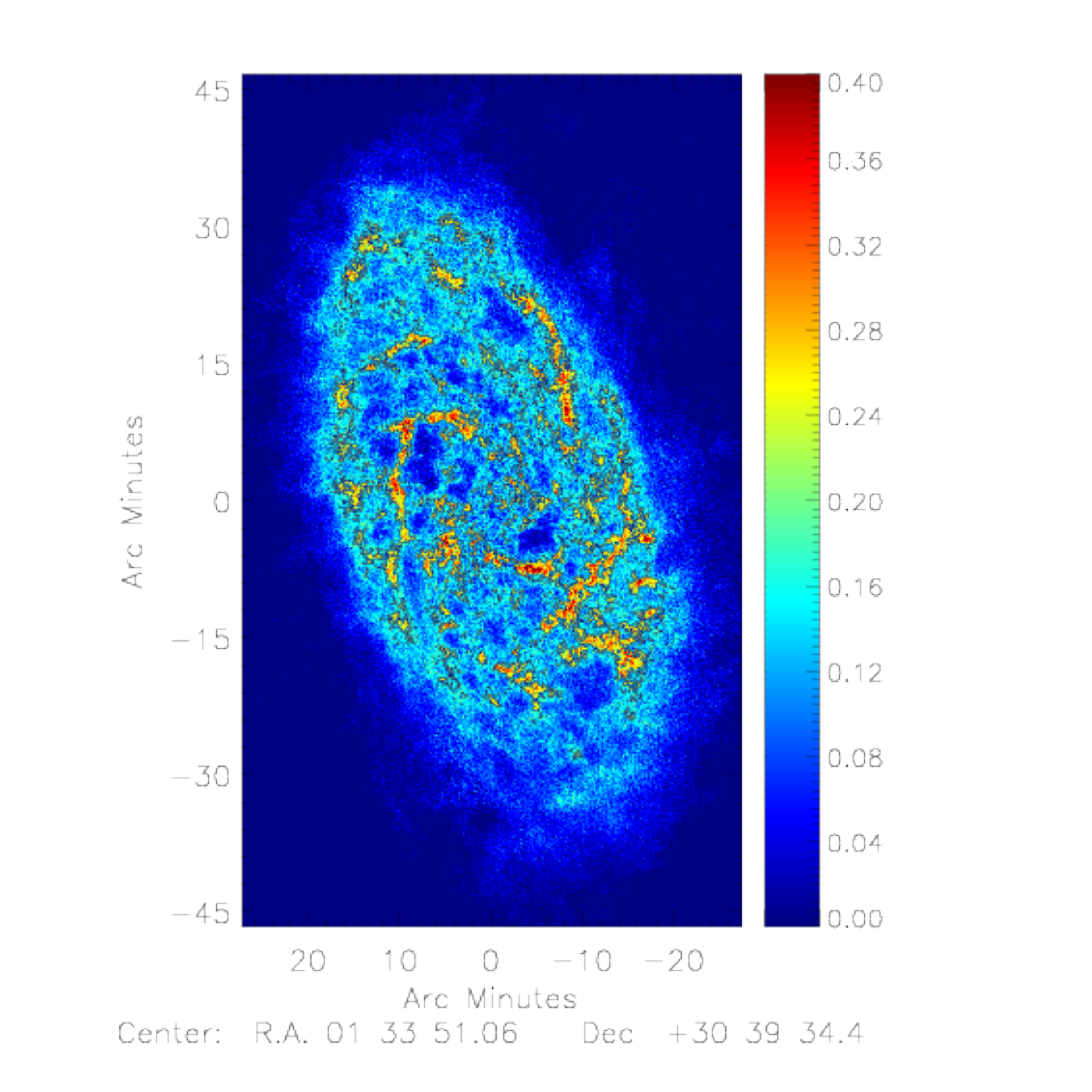}}
\caption{Velocity dispersion (moment-2) map of the 21-cm line emission  in M33  at 10~arcsec resolution ({\it left panel}).
The black contours underline the location of the bright HI filaments or arms,  they are drawn at 
0.2~Jy~beam$^{-1}$~km~$s^{-1}$ and   overlayed also on the moment-0 map  in the {\it right panel}. 
The two areas that have been masked out for the likely presence of gas infall/outflow   (the stream
towards the central region in the southern half and a smaller region above the northern arm) have
anomalously large velocity dispersions and have yellow-red  colours in the velocity dispersion map. The colour legend for the
velocity dispersion map ({\it left panel}) is in units of km~s$^{-1}$,  for the intensity map ({\it right panel})  is in  
Jy~beam$^{-1}$ km~s$^{-1}$. Coordinates of the map centers are written in the bottom part of the Figure.
}
\label{disp}
\end{figure*}

The line of sight velocity dispersion $\sigma_{21}$, defined above, is assumed to be isotropic. 
The map of the velocity dispersion of the M33 disk at 21-cm, can be an indicator of the turbulence and kinetic energy
of the HI gas, especially for nearby galaxies which can be imaged at high resolution to avoid
line broadening due to disk dynamics. We show the  moment-2 map in the left panel of
Figure~\ref{disp} using the black contours to outline the high density HI filaments or arms,
as shown by the 21-cm moment-0 map in the right panel of the same figure. 
In the dispersion map we masked out all pixels
corresponding to the two regions with large infall/outflow of gas: the stream
towards the central region just south of the nucleus, and a smaller region above the northern arm. 
In these areas the signal is a superposition of emission lines at two velocities, one corresponding
to gas in the disk  and the other being due to extraplanar gas. 

\subsection{The molecular gas and the GMCs population}

We use molecular data of the deep $^{12}$CO(J=2-1) whole-disk survey
carried out with the IRAM-30m telescope \citep{druard14}  at 10.7$"$
resolution (44~pc). From the spectral cube of this database at a spatial resolution of
12$"$ and at a spectral resolution of 2.6~km~s$^{-1}$,   a catalogue of 566 giant
molecular clouds (GMCs) has been presented by \citet{2017A&A...601A.146C}, 
following a procedure described and used for M33 by \citet{2012A&A...542A.108G}.
The main beam temperature noise  for the dataset at  12$"$  resolution
is  0.033~K in each
channel. The moment-0, or brightness map that we use  has 3~arcsec pixels and on average
$\sigma_{CO}$=0.27~K~km~s$^{-1}$, considering a typical   spectral window of
25~km~s$^{-1}$ per beam. Using a CO-to-H$_2$ conversion factor
X=$N(H_2)/I_{CO(1-0)}=4\times 10^{20}$~cm$^{-2}$/(K~km~s$^{-1}$) \citep{2017A&A...600A..27G}, twice
the galactic value, and an intrinsic line ratio $R^{2-1}_{1-0}=I_{2-1}/I_{1-0}=
0.8$  \citep{druard14}, we estimate $\sigma_{H_2}$= 1.3$\times
10^{20}$~H$_2$~cm$^{-2}$.  This is an average estimate of the map noise because
there are variations of the main beam temperature noise and of the signal width
across the disk.  To trace the PDF we consider only pixels with flux above 3 $\sigma_{CO}$ and
trace the statistical analysis down to the corresponding limiting column density,
which in term of total H atoms reads 3 $\sigma_{H} \simeq 7.8\times
10^{20}$~H~cm$^{-2}$  (i.e. $4.6\times 10^{20}$~H~cm$^{-2}$ face-on value).

Molecular clouds are classified in three broad categories: clouds without obvious
star formation (A), clouds with embedded star formation (B) and clouds with
exposed star formation (C) by  \citet{2017A&A...601A.146C}.  An on-line catalogue lists
several cloud properties such as cloud radius and luminous mass computed by
converting the total CO  line luminosity of a cloud into mass  as given by
equation (1) of \citet{2017A&A...601A.146C}. Type A,B,C clouds
have average luminous masses of 1.3, 2.6 and 3.1~$10^5$~M$_\odot$. As noticed
by \citet{2017A&A...601A.146C} and by \citet{2018A&A...612A..51B},  going from non-star-forming to
exposed star-forming clouds, the average cloud mass grows.  The total H$_2$ mass in
A,B,C-type clouds is, respectively, 0.16, 0.14 and 0.79~$10^8$~M$_\odot$.  Adding
the unclassified D-type clouds, we have a total molecular hydrogen mass  of
1.12~$10^8$~M$_\odot$ in GMCs or a total GMC gas mass (considering He and heavy
elements, i.e., taking a hydrogen fraction of 0.73~$\%$) of 1.53~$10^8$~M$_\odot$.
That is about half of the total molecular ISM mass. The rest is in lower-mass
molecular clouds (below the survey resolution limit) which we shall call ``diffuse
molecular gas'', even though some of this gas may be self-gravitating.
By selecting 630  mid-infrared (hereafter MIR) sources from the list of \citet{2011A&A...534A..96S} that
are YSCCs in the early
formation and evolutionary phases, \citet{2017A&A...601A.146C} determined that they are
strongly correlated with the GMCs (the correlation length is of order 17~pc, smaller than 
the typical GMC radius). Since age
estimates of  the YSCCs are between 3.5 and 8~Myrs, given the fractions of GMCs in each
class (i.e., presumably in each evolutionary phase)  the GMC lifetime  was
estimated to be 14.2~Myrs, with the longest phase being the C-type phase, when the
YSCCs break out of their clouds before total gas dispersal.

To sample the surface density distribution for each cloud, we can either use the
corresponding pixels identified by the cloud extraction algorithm, or use the
cloud radius r$_c$ to find the associated pixels. Although most of GMCs are rather
spherical, there a few of them which have an elongated shape and look filamentary.
We have  carried out the analysis using both selection criteria at first but found
no sensible differences in the resulting PDFs. We therefore decided to use the cloud radii
listed by \citet{2017A&A...601A.146C}  to select  pixels corresponding to each GMC. This
simplifies the procedure to find the atomic gas at and around cloud positions. The average GMC 
radius decreases from 26 to 22 to 18~pc from A- to B- to C-type clouds, even though the mass 
increases in this sequence.
 This seem to suggest that GMCs are more compact as they evolve, likely collapsing and shrinking 
 as they collect more material and begin to form stars. It is not clear however if GMCs of
type-C are more bound and dominated by gravity since the presence of young stars in these molecular
complexes  might increase the local turbulence.  It is therefore of
interest to investigate the PDFs of GMC at each evolutionary stage.

\subsection{Selected regions and total gas maps}

\begin{figure}
\centering
\includegraphics[width=9cm]{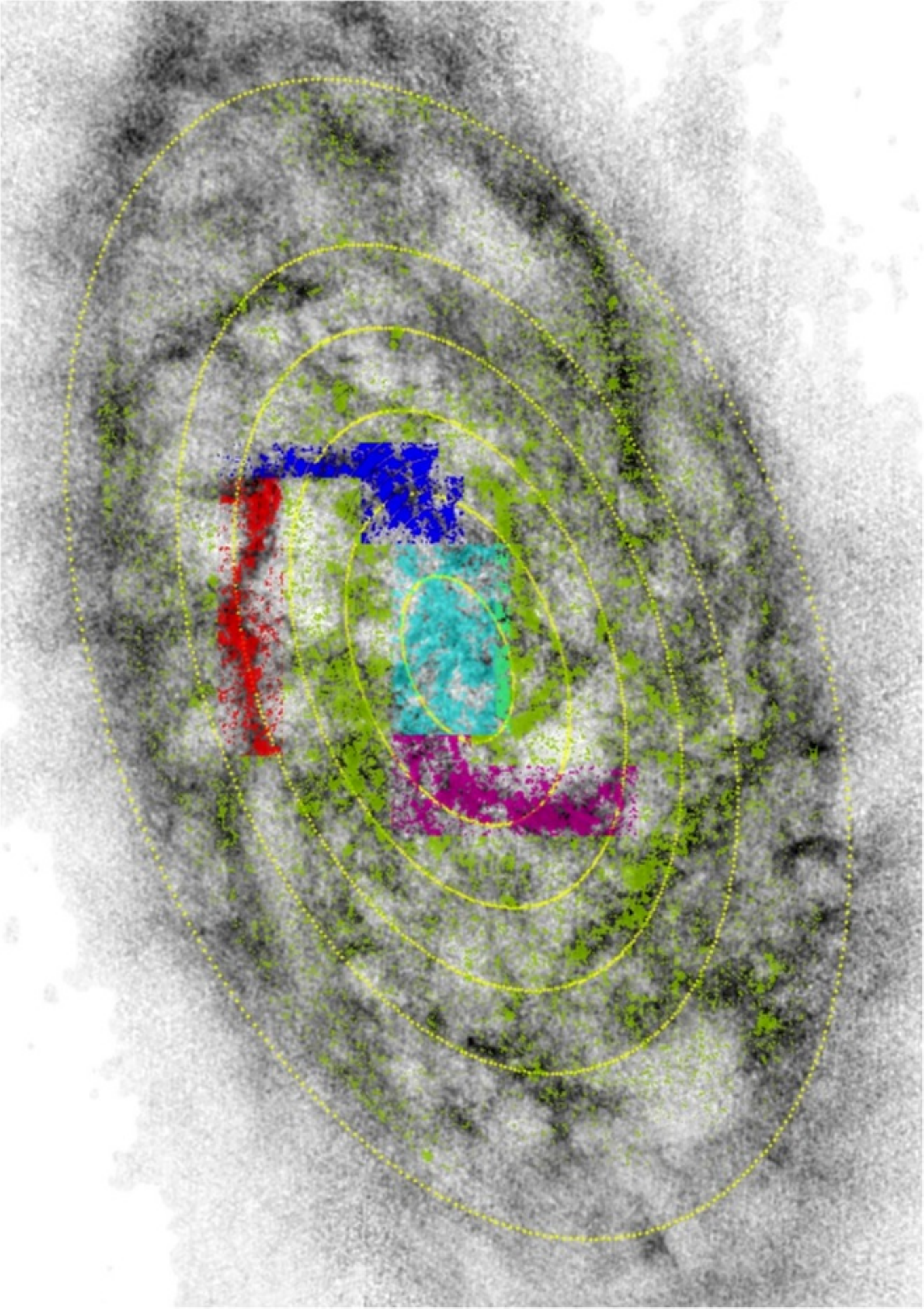}
\caption{Map of regions selected for this study. The grey background image is the 21-cm map of M33 and the 6
yellow ellipses indicate the boundary of the radial bins: 1,2,3,4,5,7~kpc.. The colored dots indicate all pixels
where CO J=2-1 integrated emission is above 3$\sigma_{CO}$ and are color coded according to the
different selected regions described in the text:  cyan for central region (CR), blue for northern arm (NA), 
magenta for southern arm (SA), red for eastern filament (EF), green for interarm (IA).
}
\label{map}
\end{figure}

Using the cold gas images we have selected four regions which might be of interest for
cloud formation and destruction and have enough pixels with CO emission above the
thresholds to ensure a  statistically significant analysis. These are: the central
region (CR) and three other regions that have  prominent HI overdensities and
non-negligible H$_2$ fractions, namely the northern arm (NA), the southern arm
(SA) and the eastern filaments (EF). The latter feature is  the HI overdensity that starts at 
NGC604 (the brightest HII region in M33 located at 9.1~arcmin to the north and 7.5~arcmin
to the east in Figure~\ref{disp}) 
and extend radially outwards (as an extension of the northern arm towards the south).  
Figure \ref{map} shows these regions on
the 21-cm emission map of M33. We mark  all pixels with CO J=2-1 integrated emission
above 3$\sigma_{CO}$ using one colour for each selected region and using the green colour
for the interarm regions (IA). We shall refer to interarm region when we analyze everything
that is not in one of the selected areas despite there are filament segments which 
are included in it. We do not have the sensitivity to select only the truly interarm ISM. 

By analyzing individual regions or radial bins separately, we minimize mixing data
relative to  gas with different scale heights and disk gravity. The southern
 arm looks less disturbed than the northern arm and it is the only arm that
 fits well with density wave models, as underlined by \cite{1980ApJS...44..319H}. This
 justifies a separate analysis of the PDFs relative to each arm.  Results for the eastern
 filament have been checked by considering also the same region without the area in
 and around NGC604. This has been done in order to make sure that it is not only the
 presence of the large star forming region  driving the detected properties.

When we analyze GMCs in selected regions we require that the center of the GMCs to
be within the boundaries of the region (not the whole GMC or its HI envelope). To
collect pixels with CO  or HI emission above the threshold associated with the
$n$-cloud, we require that the distance d$_{i,n}^{CO}$ and d$_{i,n}^{HI}$ between
the $i$-pixel center and the $n$-cloud center satisfies
\begin{equation}
d_{i,n}^{CO} \le 1.7 \times r_c^n  \ \ \  d_{i,n}^{HI} \le 2.5\times r_c^n.
\end{equation}

For gaussian shaped radial distribution of cloud brightness  we recover 90$\%$ of the 
total  cloud  molecular mass with this selection. For the HI envelope the factor 2.5 
encompasses pixels next to the GMC without appreciable CO emission. 

The total hydrogen gas map is the sum of the molecular and atomic gas surface
densities pixel by pixel. The CO J=2-1 map was aligned with the HI 21-cm map and
the HI and CO line brightnesses converted to H atom column densities. Before
adding the  molecular gas surface density to the atomic one, we masked out all
4~arcsec  wide pixels with H$_2$ column density below 2$\sigma_{H_2}$
(We mask pixels below 2 and not 3$\sigma_{H_2}$ because
 the alligned CO map with 4~arcsec  pixels has a lower noise than
the original $\sigma_{H_2}$ value. 
Given the subsolar metallicities of M33 it is likely that there is not much molecular gas
at lower surface densities because of radiative dissociation. Because the 
HI image has a better sensitivity than the CO image in tracing the hydrogen
column density,  the lower limit for  sampling the total gas column density 
will be the same as that of the HI gas.

\section{The global and local gas PDFs across the M33 disk}
\label{pdf}

PDFs of inclination-corrected gas surface densities were made by summing pixels in
logarithmic bins for HI, H$_2$, or total gas using units of H atoms cm$^{-2}$. One
set of PDFs was made for different radial intervals  and another set of PDFs was
made by summing pixels in four selected regions. In this last case, the PDFs
corresponding to three types of GMCs  and the diffuse CO are also shown (diffuse
CO is where CO emission is associated with unresolved low mass clouds or is truly
diffuse). The three GMC types are: GMC-A (containing no trace of star formation),
GMC-B (with only embedded star formation), and GMC-C (with exposed HII regions).

In order to localize deviations of the PDF from a log-normal distribution we investigate
the local slopes defined as the derivative of the PDF in every bin of column density
( logarithmically spaced). Local slopes are computed fitting a line segment through
 two adjacent bins on each side. A log-normal PDF will have a linearly decreasing
 slope while a power law PDF will have a constant slope. Fitted log-normal PDFs to the data
 are also shown to complement the analysis  for the most relevant cases. As discussed in
 the next Section and  mentioned in Section~\ref{intro}, compressible turbulence
 shapes the PDF as a log-normal  while self-gravity  gives a cloud an internal density 
 structure and a power law PDF for the  gas surface density.

\subsection{The PDF across arms, filaments, clouds, and the galaxy center}

\begin{figure*}
\centering
\includegraphics[width=17cm]{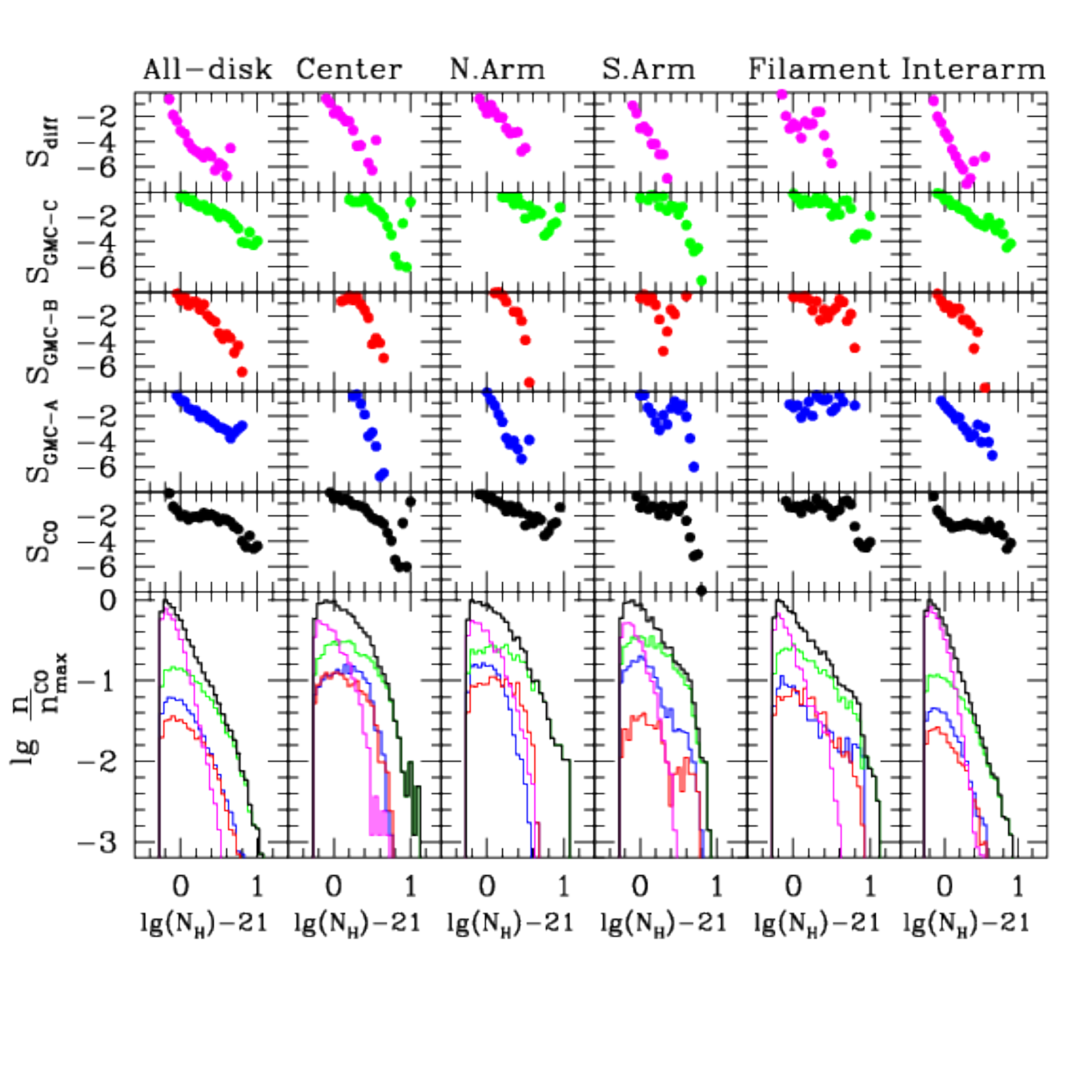}
\caption{$(Bottom\ panels)$ The PDFs of molecular gas column density in units H-atoms~cm$^{-2}$
for the total molecular gas in the disk (black heavy curves), and for the gas in
three GMC types: A (non-star forming), B (embedded star formation),
and C (exposed star formation) in blue, red, green, respectively in the on-line version.
The PDF for diffuse molecular gas  (meaning low mass clouds or truly diffuse) is marked in magenta
in the on-line version. All PDFs have been normalized to the number of pixels at the peak of
the total molecular gas distribution (n$^{CO}_{max}$) and refer to face-on values.
$(Top\ panels)$ The local slopes S in the corresponding PDFs
for the total molecular gas, S$_{CO}$, for the surface density of type A-,B-,C-GMCs,
and for the diffuse molecular gas.
The columns correspond to different regions in the galaxy: the full disk is on the
left, followed by the central region, the northern and southern spiral arms, and the
eastern filament. The combination of molecular gas from
the rest of the galaxy is labeled ``interarm'' and it is shown on the right.
}
\label{COPDF}
\end{figure*}

Figure \ref{COPDF} shows the molecular  gas PDFs: in the bottom panels for gas in the
different GMCs types and diffuse (color coded using different colours), and with thick lines for the total 
molecular gas  (regardless of GMC type or diffuse). The columns correspond to six
different regions in the galaxy, as indicated at the top. The full disk is on the
left and includes all pixels with CO brightness above 3$\sigma_{CO}$, followed by the central region, 
the northern and southern spiral arms, and
the eastern filament. The combination of molecular gas which is not in CR, NA, SA, EF
is labeled ``interarm'' (IA) and it is shown on the right. In the top 5 panels, Figure
\ref{COPDF} shows the local slopes of these PDFs: S$_{CO}$ (black dots)
are  the PDF slopes for total molecular emission,
S$_{GMC-A}$,S$_{GMC-B}$,S$_{GMC-C}$,S$_{diff}$, are the PDF slopes for molecular
gas in GMC-type A,B,C and diffuse. The log-normal parts of the PDFs  in the bottom
panels have linearly decreasing slopes while the power law parts of the PDFs have
constant slopes and in this case the point distributions in the slope panels
flatten. The panels show linearly decreasing trends in the center and in the
northern arm while for the eastern filament constant slopes are present. For the
total CO the PDFs have power law parts in the southern arm, eastern filament and
interarm regions. The southern arm and filament show also power law parts   for
individual GMC-types. Elsewhere, the PDFs are approximately log-normal, or too
irregular to detect a deviation from log-normal. 

For the total molecular gas in
the whole disk (black colors, left panel), the power law part of the PDF starts at
a column density of $\sim10^{21.1}$ cm$^{-2}$, which corresponds to about 0.7 magnitudes
of visual extinction; A$_V=0.7$~mag is obtained using the conversion factor in \cite{bohlin78} 
but it can have slightly   lower values given the metallicities of M33 \citep{2010A&A...512A..63M}.
We would like to remark that the whole disk PDF shown in Figure \ref{COPDF} is the same as that
presented by \citet{druard14}, being drawn from the same data. However, we use the slope to infer 
the column density interval in which the PDF deviates from a log-normal distribution,
instead of fitting a log-normal function, given the limited range of column density sampled.
The ascending side of the PDF, at low column densities, is not traced because of the sensitivity limits
of the survey and of molecular gas dissociation thresholds. This leaves large uncertainties in the parameters
of the best fitting log-normal functions, which in the next subsection are only shown for a qualitative comparison 
with the data. The analysis of the slopes shows a power law
behaviour which starts at slightly lower column densities than those found by \citet{druard14}
(1.3$\times 10^{21}$~H~atoms~cm$^{-2}$ versus 3.4$\times 10^{21}$~H~atoms~cm$^{-2}$).
The spatial resolution of the survey limits the analysis to N$_H < 10^{22}$~H~atoms~cm$^{-2}$
since these high column density regions tend to be small in physical size and  cover only a small portion 
of the beam.  This explains the sharp drop in the molecular gas PDF at high column density.

\begin{figure}
\includegraphics[width=10cm]{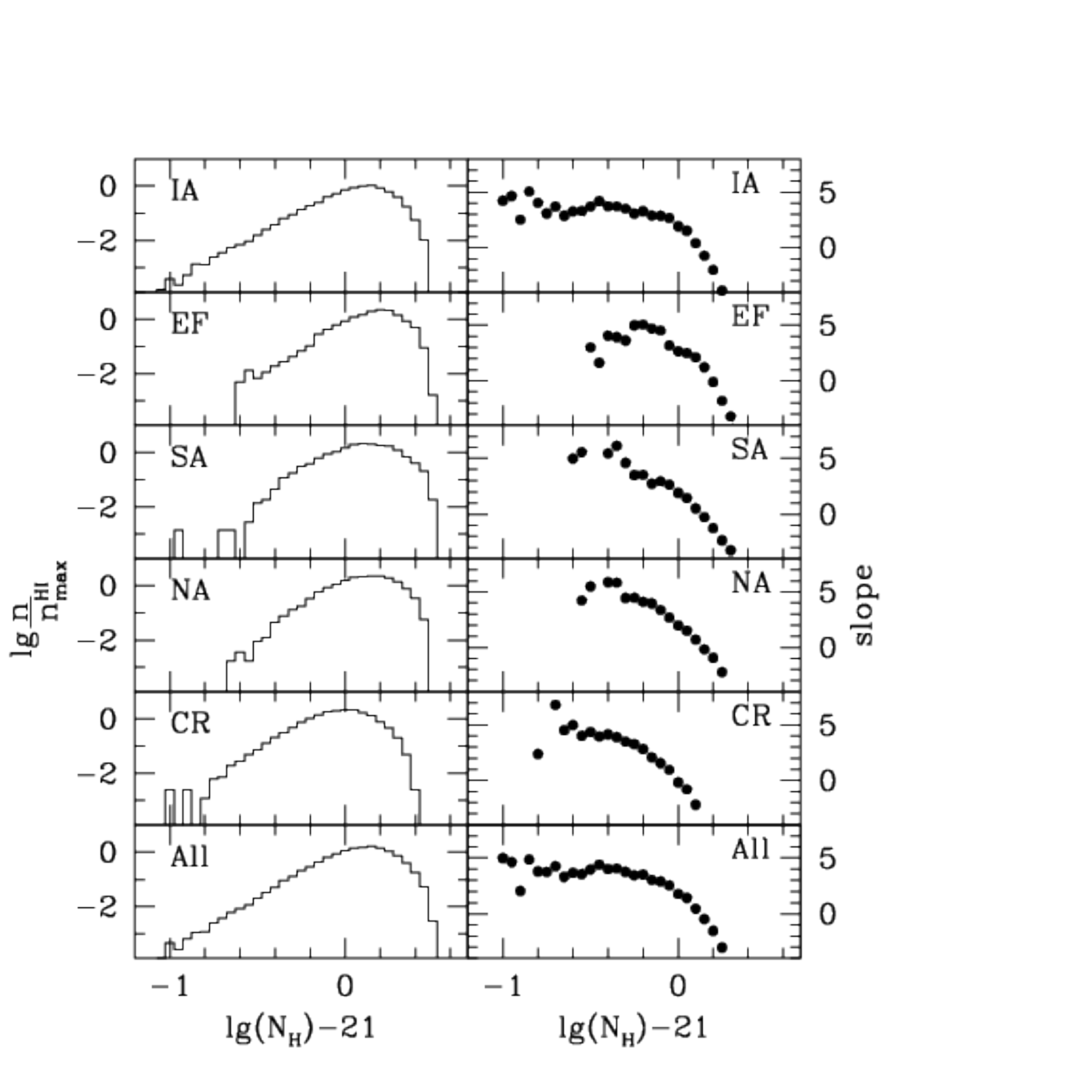}
\caption{The PDF (left panels) and the local PDF slope (right panels)
of the HI column density in and around the
population of GMCs in the whole disk of M33 (labelled All, bottom panel), in the four selected regions
(intermediate panels) and in the interarm (top panel), labelled as in Section~2.3 and Figure~\ref{map} caption.}
\label{HIPDF}
\end{figure}

In Figure \ref{HIPDF} we show the  atomic gas PDF obtained from  HI in the proximity of GMCs 
i.e. 21-cm column densities   at a distance less than 
2.5 cloud radii with respect to the cloud centers. It is evident that the PDF drops at high 
column density, above N$_H^{lim}$=2.5$\times$ 10$^{21}$~cm$^{-2}$.
In the central region the drop of the HI PDF in cloud envelopes happens at slightly lower column
densities. To interpret N$_H^{lim}$ we recall that
the molecular fraction as a function of the total gas column density has
been investigated in several galaxies and also in M33 (Figure~13 of \citet{2007A&A...473...91G}).
The gas mass fraction in molecular form increases as the gas  
surface density increases, with lower fractions found for gas at large galactocentric distances.
However, the trend of molecular gas mass fractions with atomic gas surface densities   
has not been explicitly  investigated in M33. With the sensitivity and resolution of the current surveys
we find average molecular mass fractions of order 0.5 as the HI column density approaches N$_H^{lim}$
at galactocentric distances less than 4~kpc; lower average fractions, of order 0.3, are found in the 
outermost regions. Hence N$_{HI}\simeq 2-4$ N$_{H_2}$ when the HI column density 
is of order N$_H^{lim}$, and we interpret N$_H^{lim}$ as the maximum
sustainable atomic column density  in the M33 disk. Molecules can form  where the 
atomic gas column density is lower than N$_H^{lim}$, especially in the inner disk where gas volume densities 
and dust abundances are high.  However, molecules always form  
where N$_{HI}$ is close to N$_H^{lim}$ or the total gas column density is higher than 2 N$_H^{lim}$
throughout the whole star forming disk of M33 (or A$_v > 2.4 $ if we consider metallicities lower
than about 0.4~dex than solar in  M33).
 
The HI PDF is approximately a power law at low column density.
This is particularly evident in the interarm PDF and  in the total  HI PDF (labelled All in Figure~\ref{HIPDF})
which both extend
towards lower column densities. Figure \ref{HIPDF} clearly shows deviations from
a log-normal distribution in the HI PDF of All, EF and IA PDFs. Slopes are constant
for N$_H < 5x10^{20}$~cm$^{-2}$. 
What causes the deviation from a log-normal distribution at low column
density? Numerical simulations of density fluctuations in compressible polytropic
turbulence by \citet{pvs98} suggest that at low density the PDF  can approach
a power law if the polytropic index $\gamma >1$. Values of $\gamma>1$ imply an
equation of state  where the Mach number increases with decreasing density.
Then the PDF gets broader than the isothermal case at low density, producing the
power law instead of a log-normal. This is a picture that may apply to the
interarm and other low-density regions if the voids produced by feedback or
divergent flows in a spiral arm have higher Mach numbers than the clouds.
We can exclude that the power law PDF observed for the HI at low column density is due to
noise or incomplete sampling in interferometric observations.
\citet{2016A&A...590A.104O} have in fact shown that the impact of the observational noise 
needs to be at least 10$\%$ of the PDF peak column density to affect the shape of the PDF
at column densities 5-10 times lower than the peak. But the noise level in the HI map is  about
5$\%$ of the PDF peak column density and deviations from a log-normal are detected starting  
from column densities only 2-3 times lower than the peak. Moreover,  our use of single disk data to
complement  VLA interferometric observations and achieve a complete  $uv$-plane  coverage
should avoid possible PDF distortions. 
However, before drawing any definite conclusion on what  drivers the low column density PDF,
one should consider possible resolution limits which can be relevant  if, for example, the low density 
gas were made of unresolved low mass clouds.  

\begin{figure*}
\centering
\includegraphics[width=15cm]{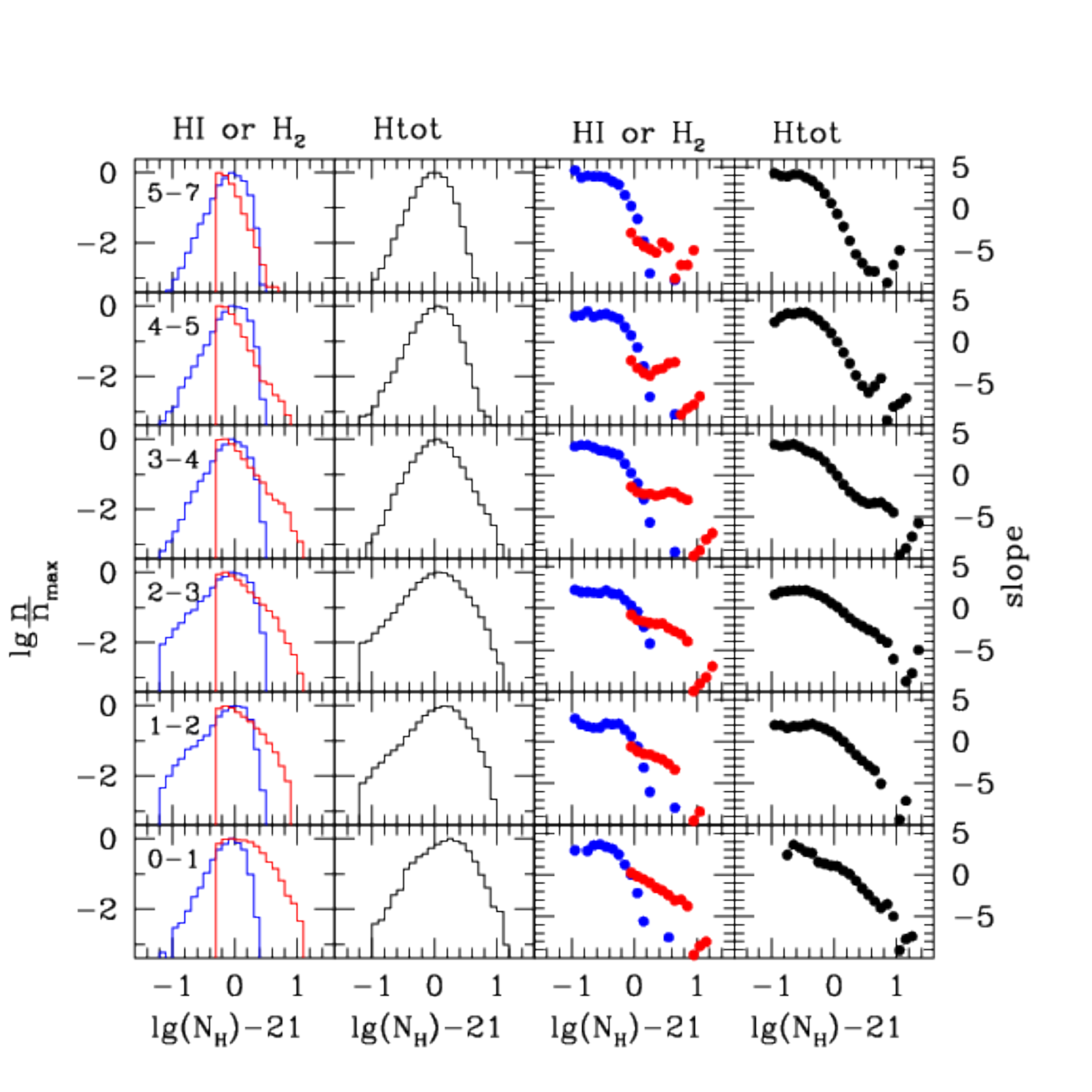}
\caption{The PDFs (left) and PDF slopes (right)
of the atomic (blue), molecular (red) and total (black)
hydrogen gas in 6 radial bins increasing
from the bottom to top and separated by 1 kpc intervals for all but the top row,
where the radial interval is 2 kpc. The label in the upper left corner of the leftmost panels
indicate the radial boundaries of each bin in kpc.
Each PDF has been normalized to their peak value.
  The 4th bin from the bottom (radius 3-4 kpc)
has a clear power-law structure in the molecules and total gas. }
\label{RPDF}
\end{figure*}

Figure \ref{RPDF} plots the PDFs for HI (blue), H$_2$ (red) and total H
(black) in the left two columns, and the slopes with the corresponding colors in
the right two columns. The six rows are different bins of galactocentric distance,
which are, from bottom to top, 0-1, 1-2, 2-3, 3-4, 4-5, and 5-7~kpc.  The 4th bin
from the bottom shows the clearest power law structure for the molecules, with index -2.2,
starting at a column density of $\sim10^{21}$ cm$^{-2}$. In this radial bin the power law is even 
present for the total gas, starting at $\sim10^{21.4}$ cm$^{-2}$, and  the PDF is
highly symmetric between the high and low density side. The large stellar cluster NGC604
and the gas in its proximity lie in this radial interval.
Approximate power laws are also present in the two outermost  radial bins
in the molecular PDFs.  Here
some caution should be used in interpreting the sharper PDF fall-off  of the molecular gas
which can be caused by dark CO i.e. by a population of compact molecular clouds, 
undetected by the CO survey, or by radial variations of the
CO-to-H$_2$ conversion factor. A higher CO-to-H$_2$ conversion factor at large
galactocentric radii would increase
the PDF at high column densities and give a more symmetric total gas PDF as well.

The atomic gas PDFs in separate radial bins, as displayed in Figure \ref{RPDF}, show similar drops  
at high column densities as  the PDFs of HI cloud envelopes in Figure \ref{HIPDF}. 
On a linear scale the HI column density is distributed smoothly around a peak
value at about 0.5 N$_H^{lim}$ with only a few pixels extending above N$_H^{lim}$.
The galaxy has no HI gas with face-on column densities above 
4$\times$ 10$^{21}$~cm$^{-2}$. We have  used the 21-cm map at a higher spatial
resolution (half of what has been normally used in this paper)
to check this finding, but no  significant variations have been detected. 
We cannot exclude that there are localized opacity corrections to the  HI column density above
N$_H^{lim}$ \citep{2009ApJ...695..937B,2012ApJ...749...87B}.
As in Figure \ref{HIPDF}, the HI PDFs drop more slowly than a log-normal at 
low column densities and they are close to a
power law. The power law slopes are between $\sim2$ and $\sim4$ outward of 2~kpc,
increasing towards larger galactocentric radii.

\subsection{PDF widths and the 21-cm line velocity dispersions}

\begin{figure}
\centering
\includegraphics[width=8cm]{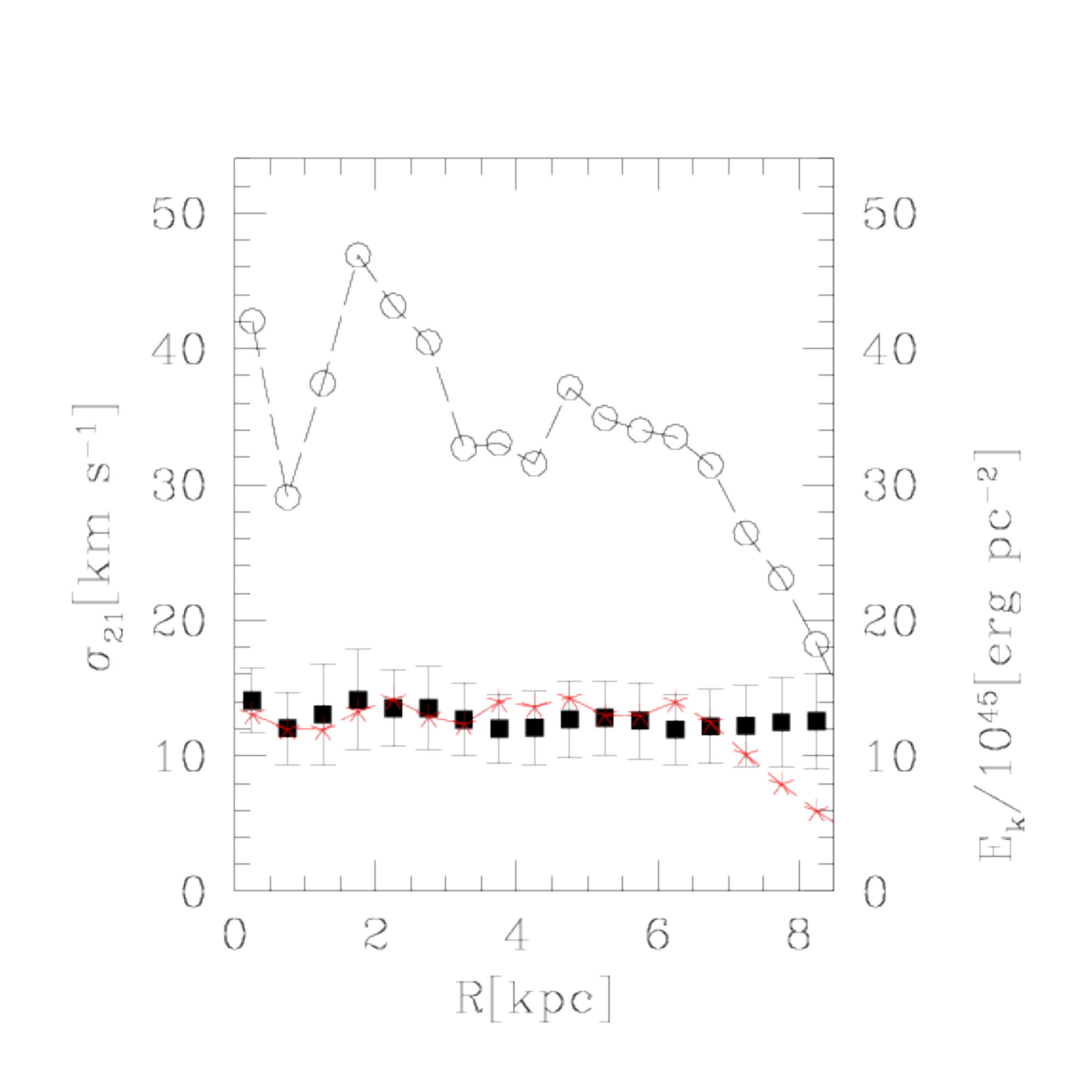}
\caption{The radial average values of the 21-cm velocity dispersion and their standard deviations
are shown with filled square symbols with errorbars.
The open circles  indicate the average energy per square parsecs in the same radial bins; energy values have
been normalized to 10$^{45}$~ergs~pc$^{-2}$. The  asterisks connected by a dashed line (in red in the
on-line version) indicate radial averages of the atomic gas surface density in solar masses per square parsecs.
}
\label{kin}
\end{figure}

The widths of total gas PDFs get narrower going from 1~kpc to large galactocentric radii.
If  the possible presence of  dark CO keeps uncertain  the widths of the molecular PDFs at
large radii, this radial trend can be inferred also from the low column density side of the PDF, 
dominated by the atomic gas.
However, there is no radial trend for the HI velocity dispersion, as shown by 
the 21-cm  moment-2 map in Figure~\ref{disp}. In Figure~\ref{kin} the filled
squares indicate radial averages of the 21-cm velocity dispersion, $\sigma_{21}$,
with their standard deviations. The flatness of the distribution is remarkable and the
resolution of the 21-cm survey for this galaxy is excellent to avoid line broadening 
due to large scale motion in the disk, such as a  rising  rotation curve in the center or spiral arm
dynamical perturbations. For  several galaxies the observed 21-cm line velocity dispersion decreases towards
large galactocentric radii \citep{2009AJ....137.4424T}. We would like to point out   that if
we lower the spatial resolution of the HI survey by a factor 10 we also detect a decrease of the
velocity dispersion from a central value of 21~km~s$^{-1}$  to  about 14~km~s$^{-1}$ at the edge of the 
star forming disk. The resolution and sensitivity of the 21-cm dataset used in this paper, 
together with the regular morphology and low inclination of this nearby spiral galaxy, seems 
appropriate for using the 21-cm line broadening
as tracer of turbulent motion. There are  some
variations in the HI dispersion across the disk, and this can be seen in
Figure~\ref{disp}. 
 
 In Figure~\ref{kin} we use the
open symbols to plot radial averages of the kinetic energy per unit area estimated
as $E_k=1.5 \sigma_{21}^2 \Sigma_{HI}$ in units of 10$^{45}$~ergs/pc$^2$.  Radial
averages of the velocity dispersion  and of the HI surface densities are about
constant across the star forming disk \citep{2014A&A...572A..23C}, but since their
small variations are correlated in the inner disk,  variations in the
kinetic energy per unit area  are more evident, even though  they are less
than a factor 2. The energy density $E_k$ peaks around 2~kpc  due to the presence of the southern and
northern arms where the kinetic energy  density in the atomic gas is about
5~10$^{46}$~ergs~pc$^{-2}$. The energy  density decreases beyond 3~kpc and
drops at the edge of the star forming disk at about 7~kpc  because  of the low gas
surface density there.  Between the four selected regions, the interarm  and the
eastern filament regions have the lowest  kinetic energy densities for the
atomic phase. The eastern filament is an HI overdensity but the HI velocity dispersion is
lower by about 2~km~s$^{-1}$ than in the arms.

Values of the velocity dispersion are well above 6-8~km~s$^{-1}$,  which is the
expected thermal width for the warm phase of the ISM, hence turbulent velocities
dominate across the whole disk. The decrease of  PDF width for total and atomic
hydrogen at large galactocentric radii  could then be due to a decrease in the
Mach number (which is of order unity). 
Because the velocity dispersion is constant, this would imply that
the average thermal temperature of the gas increases with radius, possibly as a
result of a transition to a higher fraction of HI in the warm phase as the stellar
surface density, and hence the disk pressure, decreases.

\begin{figure}
\centering
\includegraphics[width=11cm]{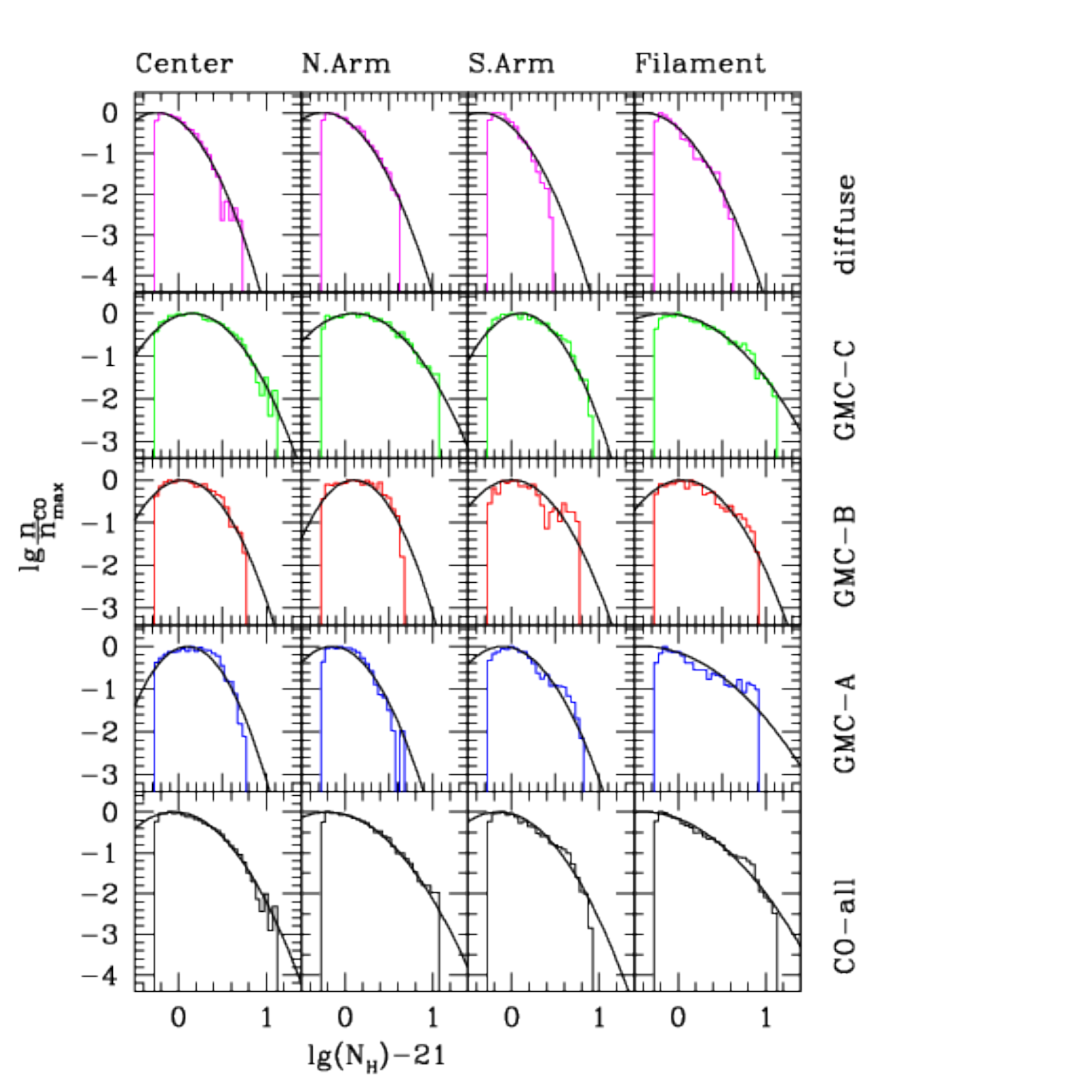}
\caption{The best fitting log-normal functions to the PDF of total CO, of CO in GMCs of type A,B,C and diffuse
in the four selected regions of the M33 disk. Each PDF is normalized to its maximum value.}
\label{fits}
\end{figure}

In Figure~\ref{fits} we show the best fitting log-normal functions to the molecular gas PDFs
in the selected regions. It is clear that log-normal functions fit better the data
in the center and northern arm region than in the southern arm and eastern
filament region. The peak  and the width  of log-normal functions have been
considered  free parameters in the intervals [$-0.4$, 0.4] and [0, 1] respectively  
(uncertainties have been estimated using Poisson statistics).
Even though the significance of the deviations are larger for the all-disk and
interarm distribution, we show the four regions here to underline some  
differences  between their PDFs.  For example in the regions where the PDFs  
are truly log-normal, namely   the center and northern arm,    
the GMCs of type C show clearly a wider log-normal distribution than  GMCs of types A and B,
likely because the newly born massive stars  in type C GMCs increase the turbulent
motion.

\subsection{Interarm and gravity dominated filaments}

In this Section we  investigate some details of the power law parts of the   molecular gas  PDFs across the
disk  that has been shown in Figure~\ref{COPDF}. 
The all-disk, the southern arm, eastern filament and interarm all show
a molecular gas PDF which deviates from a log normal at N$_H>10^{21}$~cm$^{-2}$ and it closely
resembles a power law.
To prove that the power law behavior of  the molecular gas PDF along the eastern filaments is
a property  of the gas along the filament and not a result of the gas around the bright stellar 
cluster NGC 604 located at one extreme,  we have also carried out the analysis 
excluding the pixels around NGC 604 from the analysis  but the same conclusions apply.
 
\begin{figure}
\centering
\includegraphics[width=9cm]{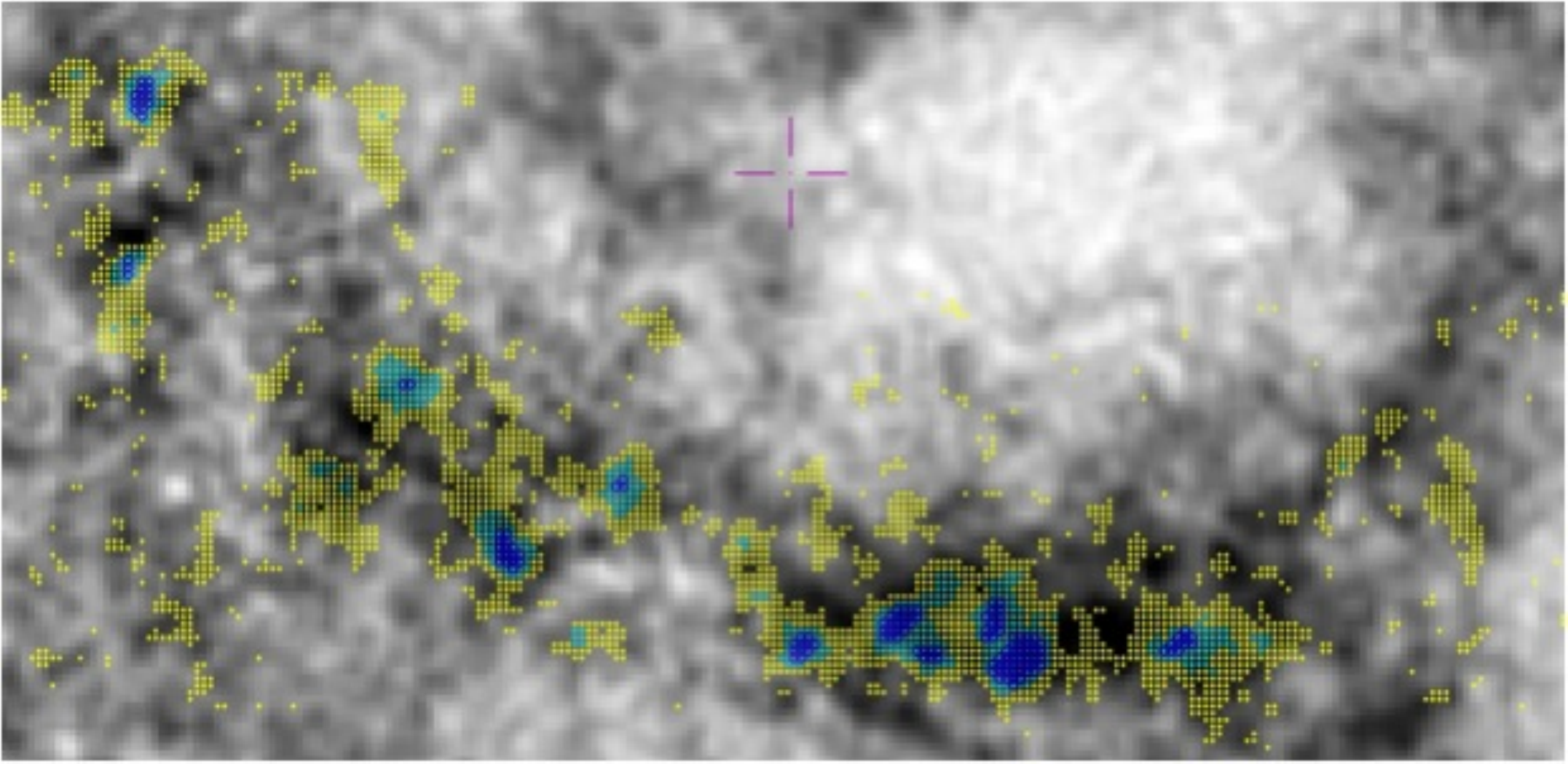}
\caption{The image of the southern arm with the HI map in the background and  colours underlining the  CO core-envelope
cloud structure which follow a  power law PDF. For the  CO emission we plot:
in yellow all pixels with molecular surface densities
10$^{21}<$N$_H<2.5\times 10^{21}$~cm$^{-2}$, in cyan all pixels that have 2.5$\times 10^{21}<$N$_H<4\times 10^{21}$~cm$^{-2}$
and finally in blue the core of GMCs with  N$_H>4\times 10^{21}$~cm$^{-2}$}.
\label{core}
\end{figure}

The power law PDF for molecules in the southern arm or in the eastern filament 
could be the result of self-gravity in GMCs,
which gives them a power law internal density profile
\citep{klessen00,vaz08,kritsuk11,2014ApJ...781...91G}.   A similar interpretation has been 
suggested by \citet{druard14} when they detected an excess in the galaxy-wide PDF at high column 
densities. To better investigate this possibility, 
in Figure~\ref{core} we color the different parts of the molecular
power law to see where they arise in the southern arm. We display the HI map of the
southern arm in the background and underline in yellow all pixels with molecular
densities that have 10$^{21}<$N$_H<2.5\times 10^{21}$~cm$^{-2}$, and in cyan all
pixels that have 2.5$\times$ 10$^{21}<$N$_H<4\times 10^{21}$~cm$^{-2}$ and finally
in blue where  N$_H>4\times 10^{21}$~cm$^{-2}$. 
The blue regions tend to be in the cores of the yellow and cyan regions, indicating that 
the power law PDF results from a power law internal structure inside GMCs. 
This is consistent with the model where cloud self-gravity makes the power law PDF in these regions.
Most of the yellow
regions with no cyan or blue pixels in the center are type-A GMCs which are
located in the inner part of the arm. These pristine clouds  presumably get
compressed and accrete mass as they cross the arm, developing a high density core
that fragments and form stars changing their status to type-C. We estimate that about 70~$\%$ of the
molecular gas in the southern arm and  eastern filament has column densities
above 10$^{21}$~cm$^{-2}$  where it deviates from a log-normal distribution.
At the resolution of our survey the slope or power law index can be as low as
$-1.2$ to $-1.5$ for CO in the eastern filament or southern arm while it is $-2$
for the overall CO PDF and $-2.8$ for interarms.

Power law PDFs are also observed for galactic GMCs and they have similar internal density gradients. 
For example \cite{schneider15c} determined power  law PDF tails above 4 to 5 magnitudes of visual extinction 
for four local molecular clouds, using dust emission maps, and showed that the power law slopes in the PDF corresponded 
to the slopes of the density distributions inside the clouds, as expected for self-gravitating clouds. 
\cite{schneider15a} measured PDFs for four infrared dark clouds using dust emission and found them to be power-law 
throughout, regardless of their internal star formation rates, with a transition to log-normal in the lower column density 
regions surrounding the dark clouds. Cloud core infall was also observed in CO, demonstrating that the power law 
structure for these clouds is from self-gravity. 
Evidently, power-law PDFs are common and galactic studies show that the origin of those power laws can often be traced 
to power-law internal density structures inside individual clouds. Some of the excess dense gas can be from local compression too, 
as demonstrated for particular regions near bright nebulae \citep{schneider12} .

The power law index increases when we average data from regions where some gas follows 
a power law and some gas follows a log-normal distribution.  However, we would like to 
point out that a power law PDF for column density might have  another explanation 
that applies to the interarm region. Here the presence of small unresolved 
cloudlets implies that the gas that is observed in each resolution element can be described as 
a random collection of individual clouds which follow a power law mass
distribution function, $dP/dM\propto M^{-\kappa}$. 
For $\kappa\sim1.5$ \citep[e.g.,][]{solomon87}, the distribution function
of column density is a power law with a slope of about $-1.5$ on a log-log
plot if the average gas density  is about constant and cloud surface density and
size scale as M$^{1/3}$.  Steeper mass function slopes   such as $\kappa\sim1.8$ have been
observed for molecular clouds in our Galaxy or in external galaxies \citep{heithausen98,2005PASP..117.1403R}
and these would be better in agreement with the observed slope of $-2.8$ for the power law of the interarm PDF
if clouds are unresolved. More recently \citet{2018A&A...612A..51B},have shown that in M33 the slope of 
the cloud mass spectrum steepens from $\kappa=1.4$ to $\kappa=1.9$ going radially outward. 
The mass distribution of star clusters in interarms  suggests also a steeper cloud mass spectrum ($\kappa\sim2$)
in these regions \citep{2018MNRAS.477.1683M}.   
In the interarm, there appears to be no correspondence
between different parts of the PDF power law and different parts of individual clouds,
such as a core-envelope structure. Also, samples of each
cloud type in the interarm seem to have   log-normal PDF, representing
turbulent conditions.  Still the PDF for all of the clouds is close to a
power-law, and that may be a stochastic effect from cloud sampling when the cloud
size is smaller than the beam.   A comprehensive
analysis of theoretical predictions for the observed PDF of HI and H$_2$ clouds 
below the spatial resolution limits of the surveys needs further investigation and  will be addressed 
in forthcoming paper. 
 
\section{Discussion}
\label{models}

 Figures \ref{COPDF} and \ref{RPDF} indicate that the gas with high surface
density in M33 has a PDF that is log-normal in the central regions and northern
arm, and power-law for the molecular emission in the southern arm and filaments
and in the interarm regions. The molecular gas has a power law PDF in general at
mid-radii, and even the total gas PDF is somewhat power law at 3 to 4 kpc radius.
The total gas has more of a log-normal PDF,  presumably because it includes
both the atomic and  molecular phases with the latter being dominant at high
column densities.
The question we set out to address in this paper is where and how the transition
from weakly self-gravitating gas to strongly self-gravitating clouds occurs.  The
most direct interpretation of the data presented in the previous Section is that
this transition occurs at the same time as the gas converts from atoms to
molecules, because the signature of self-gravity, the power law PDF, is primarily
in the molecules at column densities close to $\Sigma_{lim}$. However, this is not
necessarily the case. It could be that much of the HI is also part of the
power-law cloud structure, perhaps even an extension of the same power law, in a
shielding layer around the molecules, and we cannot see that because the turbulent
compressions in the HI are much larger than the radial density range.

The power law PDF is usually interpreted as the result of a turbulent cloud with a
radial gradient in the average density \citep{klessen00,vaz08,kritsuk11}. The
turbulence presumably gives each small region a log-normal PDF in volume density,
but the centroid density of that PDF varies with radius in proportion to the
average density (for a constant compressive Mach number). The sum of all the local
log-normals in a whole cloud becomes a power law, overwhelming the local
log-normals, when the total density range from the radial gradient exceeds the
fluctuating density range in each local piece of turbulence \citep{elmegreen11}.
This situation is fairly easy to achieve for the molecular gas because the density
range in the radial gradient can be large, a factor of 100 or more, and that range
exceeds the local compressions from turbulence at a modest Mach number.  For an HI
envelope the radial density range can be much smaller, perhaps only a factor of 10
from the cloud edge where the turbulent pressure equals the ambient pressure to
the base of the envelope where molecules appear.   If the density variations from
turbulent compressions in the HI envelop exceed the average density variation over
radius, then the power law component of the HI PDF may not be seen.  The HI PDF at
higher than average density mostly shows the sudden drop from the conversion to
molecules, so the radial structure of the gas cannot be determined this way.

This model where turbulent fluctuations dominate density gradients in the
atomic envelopes of clouds explains why the power law parts of the PDFs
generally disappear in the total hydrogen gas, which  approximately follows
a log-normal with smaller widths at larger galactocentric radii. The HI and total
gas are close to a log-normal form around the peak, even though the molecular
parts of the clouds and perhaps each entire cloud has a power-law radial profile
in density. The PDF slope levels off to a constant value at the high density, 
molecule-dominated end, where the  internal power law density structure of
the cloud dominates the turbulent density fluctuations in the molecules.  For
small clouds, the transition from atoms to molecules happens at scales that are
not sampled by the 50 pc resolution of the present survey.

Power law PDFs at high column density can also appear even without self-gravity if
the polytropic index, $\gamma$, is less than 1. According to \cite{pvs98}, low
$\gamma$ leads to an increasing Mach number at higher density because the thermal
temperature drops, and this causes higher density compressions than in an
isothermal gas. The result is a broadening of the PDF at high density and a power
law structure for the case they considered.  However, cloud temperature gradients
might reverse when star formation is taking place in the cloud core, like for C-type GMCs.  
 While this $\gamma$ effect may be
present to some extent, numerical simulations of cloud assembly and collapse also
show power laws just from self-gravity.
For example, the PDF of the eastern filament (Fig. \ref{COPDF}) resembles the PDF
in Figure~2 of \citet{2011MNRAS.416.1436B} (panels relative to 13-14 Myrs) 
for numerical simulations of  cloud formation and evolution  
by converging flows by \citet{2008ApJ...689..290H}. In this model GMCs assembly 
through thermal and gravity-driven instabilities in the atomic gas and they start collapsing
around 10~Myrs with self-gravity driving departures of the PDF from a log-normal.
Indeed, the eastern filament lies in an
atomic-dominated region, similar to that of the numerical simulation, and the
resolution and its inflow Mach number are also similar to that of our data (i.e., the Mach
number is 1.5 resulting from a velocity dispersion of $\sigma_{21}=11$~km~s$^{-1}$
in warm HI gas).

The shape of the column density PDF on large scales may also depend on the
relative times spent in the turbulence-dominated phase and in the gravity-dominated
phase.  If gravitational contraction to a power-law density structure occurs
during cloud formation, and feedback-driven turbulence occurs during cloud
disruption, then regions with a gradual formation and a fast disruption should
have a higher fraction of the mass in the pre- and early star formation phases
when the power-law PDF is still apparent.  Whether these phases show up in
molecules or atoms depends also on the ease of forming molecules. In the central
regions of galaxies where the pressure is high, a high fraction of the ISM can be
molecular, including even some non-gravitating clouds. This distinction may be
relevant to our observation that regions close to the galaxy center have a
molecular PDF that looks like a log-normal, while filaments further out have a
molecular PDF that looks more like a power law. Near the center, the molecular PDF
apparently includes more of the turbulent ISM and in that sense is like the
atomic-gas PDF further out. The high pressure   in the central regions  increases
the column density threshold for strong gravity, making this regime still undetectable 
at the resolution of the actual surveys. The filaments further out, on the other hand, tend to form
molecules at the same time as the gas becomes self-gravitating, so the molecular
PDF shows the power law density structure in that gas. 

For the center and northern arm we notice that type-C GMCs show a wider PDF than
type-A or B GMCs with a possible sign of gravity-driven power laws at very high
column densities, of order 10$^{22}$~cm$^{-2}$. For the southern arm, departures
from a log-normal are present for all 3 types of GMCs, which share similar widths,
for N$_H\simeq 3\times 10^{21}$~cm$^{-2}$. Clearly in these regions the power law
regime driven by gravity is a short lived phase, prior to the birth of stars that
stir again the GMCs, and molecular hydrogen forms rapidly without self-gravity
\citep{2008MNRAS.389.1097D}.

What needs still to be investigated is the power law detected in the interarm PDF.
The slopes are steeper there than in individual over-density regions, and the PDFs
of individual GMC classes do not show strong deviations from a log-normal
distribution. The number of pixels with high column density is low in these
regions, so most for the detected CO does not show the core-halo cloud structure
like in the southern arm (Figure~\ref{core}). It might be that in the interarm
regions, heating is more effective in warming up the outer envelopes of clouds,
which are also of lower mass, so that we have  $\gamma < 1$ without strong
gravity. Or, for smaller mass clouds, the resolution of the survey is not enough
to measure  surface densities and  detect the  self-gravitating parts of individual clouds.

\section{Summary}
\label{sum}

We examined in detail the probability distribution functions of the atomic and molecular ISM of M33
at a resolution of about 50 pc on a cloud-by-cloud basis and in  the overall gas  distribution
in selected disk areas, to possibly unveil fingerprints of self-gravitating gas at these spatial scales.
In particular we focused on spatially resolved GMC complexes and on
sensitive high resolution atomic gas imaging to sample different regions across the star forming disk.
As shown by \citet{2018MNRAS.tmp.1241D}  gaseous spiral features, i.e. filamentary
overdensities in the atomic gas, form as a result of gravitational instabilities
in the  disk of M33.  At the same time, feedback must take place in order to form 
large cavities between the filaments, which can break  as  shock fronts propagates
in the ISM.
As the galactocentric distance increases, the rate of star formation per unit
area, as well as the gravity due to the stellar surface density, decreases. It is
not clear a priori where and if recycled turbulent gas driven by shocks in the ISM
makes a fast transition to a gravity-driven regime, or else if the disk
dynamical instabilities trigger H$_2$ formation and new star formation episodes. A
first analysis of the PDF across the M33 disk, shown in this paper, suggest that
both regimes can actually take place. 
We have sampled different regions to possibly trace
non-gravitating turbulent complexes as well as gravity bound clouds in a
statistical way using high resolution all-disk survey of the ISM in M33.
We summarize below the main results:

 \begin{itemize}

 \item The face-on atomic gas column density distribution drops  
     beyond  N$_H^{lim} = 2.5 \times 10^{21}$~cm$^{-2}$ at all galactocentric radii;
     no HI is detected at N$_H > 2 $N$_H^{lim}$. The
     peak of the atomic gas PDF is at about 10$^{21}$~H~cm$^{-2}$ (or about 0.5 N$_H^{lim}$), with a slight
     increase of this value if we sample the envelope of molecular complexes.
     Around the PDF peak the HI here has a log-normal distribution. The slope of the PDF at low
     column densities becomes constant with a power-law structure, and this
     might be connected with the presence of cavities in the interarms where the
     warm/cold interface takes place, particularly if $\gamma >1$.

 \item The width of the HI PDF has its maximum value at a distance of around
     2-3~kpc from the center, likely due to the presence the two main arms. Here
     the kinetic energy density in the atomic gas has its maximum value. The
     dispersion of the HI gas varies by about a factor 2 across the disk but
     radial averages show a constant trend with values of order 13~km~s$^{-1}$ and a
     radial decrease of only 1~km~s$^{-1}$ beyond 4~kpc.  This constant trend
     suggest a ubiquitous presence of turbulence. The smaller width
     of the PDF at large galactocentric radii can be explained as a decrease 
     of the Mach number due to disk flaring and to warm HI filling the ISM.  

  \item The PDF of the molecular gas, as traced by CO, follows a log-normal in the
      central region. This applies also to the gas associated with GMCs at
      various evolutionary stages. The molecular PDF of the
      northern arm looks very similar to that of the central region. In the
      southern arm instead a power law distribution is detected  around and
      beyond N$_H^{lim}$ likely due to gravity shaping the structure of GMCs  at
      all evolutionary stages, prior to and during star formation.

 \item The PDF of the molecular hydrogen in the outermost filament that is rich
     in CO shows  a power law with indexes between -1 and -1.5. This is
     detected also in all the evolutionary phases of the GMCs which presumably assemble from 
     atomic gas through thermal and gravitationally-driven dynamical instabilities and
     turbulence is less effective in shaping the cloud structure. There is a correspondence 
     between different parts of the PDF power law and different parts of individual clouds, such
     as a core-envelope structure, for GMCs located in the outer filament and in the southern arm.

 \end{itemize}

The analysis presented in this paper suggests that on intermediate
scales  self-gravity plays an early role in molecular cloud assembly and dominates
the internal cloud structure in regions that are free of excessive feedback.
The interpretation of the PDF shape for unresolved clouds in the diffuse molecular gas or interarm, 
which shows deviations from a log-normal distribution, requires some further investigation which
will be presented elsewhere.

\end{document}